\documentclass[runningheads]{llncs}
\usepackage[T1]{fontenc}
\usepackage{amsmath}
\usepackage{amsfonts}
\usepackage{graphicx}
\usepackage{enumitem}
\usepackage{algorithm}
\usepackage{soul}
\usepackage{algpseudocode}
\usepackage{tabularx}
\usepackage{threeparttable}
\usepackage{longtable}
\usepackage{booktabs}
\usepackage{multirow}
\usepackage{multicol}
\usepackage{makecell}
\usepackage{hyperref}

\usepackage{orcidlink}
\usepackage{marvosym}

\usepackage{color}

\begin{document}
\title{Towards Privacy-Preserving Federated Learning using Hybrid Homomorphic Encryption}
\titlerunning{Towards Privacy-Preserving FL using HHE}
%
\author{Ivan Costa\textsuperscript{({\footnotesize\Letter})}{\large\orcidlink{0009-0009-8480-9352}}
\and 
Pedro Correia\orcidlink{0009-0009-4816-1755}
\and
Ivone Amorim\orcidlink{0000-0001-6102-6165}
\and
Eva Maia\orcidlink{0000-0002-8075-531X}
\and
Isabel Praça\orcidlink{0000-0002-2519-9859}}


\authorrunning{I. Costa et al.}
%
\institute{GECAD, ISEP, Polytechnic of Porto, 4249-015 Porto, Portugal \\
\email{ivcsi@isep.ipp.pt}}
%
\maketitle              
\begin{abstract}
Federated Learning (FL) enables collaborative training while keeping sensitive data on clients' devices, but local model updates can still leak private information. Hybrid Homomorphic Encryption (HHE) has recently been applied to FL to mitigate client overhead while preserving privacy. However, existing HHE-FL systems rely on a single homomorphic key pair shared across all clients, which forces them to assume an unrealistically weak threat model: if a client misbehaves or intercepts another’s traffic, private updates can be exposed.
We eliminate this weakness by integrating two alternative key-protection mechanisms into the HHE-FL workflow. The first is masking, where client keys are blinded before homomorphic encryption and later unblinded homomorphically by the server. The second is RSA encapsulation, where homomorphically encrypted keys are additionally wrapped under the server’s RSA public key. These countermeasures prevent key misuse by other clients and extend HHE-FL security to adversarial settings with malicious participants.
We implement both approaches on top of the Flower framework using the PASTA/BFV HHE scheme and evaluate them on the MNIST dataset with 12 clients. Results show that both mechanisms preserve model accuracy while adding minimal overhead: masking incurs negligible cost, and RSA encapsulation introduces only modest runtime and communication overhead.

\end{abstract}

\section{Introduction}

Federated learning~(FL) is a distributed machine learning approach that allows a group of clients, such as Internet of Things (IoT) devices or healthcare entities, to collaboratively train a global model without sharing their sensitive data. In FL systems, clients train local models using their own data and are required to share only the model parameters. In a centralized FL system, the local models are sent to a server that is responsible for coordinating the training process by aggregating the received updates to form an improved global model. Once updated, the global model parameters are sent back to the clients, allowing them to continue local training~\cite{mcmahan2023communicationefficientlearning}. On the other hand, decentralized FL systems do not rely on a central server, and clients coordinate this process between them.
In either case, raw data never leaves clients' devices, which is the reason why FL is often described as ``privacy-preserving''~\cite{aledhariFederatedLearningSurvey2020}. 
However, recent studies have demonstrated that local model updates can still leak private information, enabling adversarial techniques such as gradient-based data reconstruction attacks, membership inference attacks, and property inference attacks~\cite{10089719,10.1007/978-981-97-5603-2_36,NIU2024404,Yu2023}.
Various techniques have been applied to overcome this problem, such as Differential Privacy~(DP), Homomorphic Encryption~(HE), and Multi-Party Computation~(MPC)~\cite{stanSecureFederatedLearning2022}.
However, these techniques also bring their own limitations:
MPC is not appropriate for resource-constrained settings with several clients, DP adds noise to the training process, and the main problems of HE, namely ciphertext expansion and noise growth, make it impractical for real world use cases~\cite{zhangHomomorphicEncryptionBasedPrivacyPreserving2023,jin2023fedmlhe}.
Nonetheless, the property of computing over encrypted data, inherent to HE, is extremely valuable to ensure privacy-preserving settings.
This has motivated the research for methods that employ HE more efficiently.

Recently, Hybrid Homomorphic Encryption (HHE) has emerged as a promising approach, combining a lightweight symmetric cipher with a HE scheme~\cite{abdinasibfarHHELandExploring2025,dobraunigPastaCaseHybrid2021}. In this type of scheme, clients encrypt data symmetrically and send both the ciphertext and the key encrypted under HE to the server. The server homomorphically evaluates the symmetric decryption circuit to transform the symmetric ciphertext into a homomorphic one, enabling computations on encrypted data without exposing sensitive information. This reduction in client-side cost makes HHE especially attractive for IoT and other resource-constrained devices.
The application of HHE in FL is still recent, and to the best of our knowledge only two works have proposed concrete systems in this direction. Correia et al.~\cite{correia2025} introduced the first HHE-FL system, based on the combination of the PASTA cipher and the BFV scheme. Nguyen et al.~\cite{michalas_FL} later proposed a similar design using the Rubato cipher together with FV and CKKS, relying on a trusted key dealer to reduce client responsibilities. While these works demonstrate the practicality of HHE-FL, they share a critical weakness: all clients rely on the same homomorphic key pair. As a result, their security depends on an unrealistically weak threat model in which no client acts maliciously. In practice, if one client intercepts another’s communication, it could recover private model updates.

Our work addresses this fundamental liability. We propose two alternative key-protection mechanisms that eliminate the shared-key weakness, thereby extending HHE-FL security beyond the honest-but-curious assumption to settings with malicious clients. Concretely, our contributions are:

\begin{enumerate}
    \item Identification and elimination of the shared-key liability present in prior HHE-FL systems, which made prior systems vulnerable to malicious clients intercepting communications.

    \item Integration of two alternative countermeasures into the HHE-FL workflow: (i) a masking mechanism, where client keys are blinded before homomorphic encryption and unblinded by the server; and (ii) RSA encapsulation, where homomorphically encrypted keys are additionally wrapped under the server’s RSA public key.

    \item Security analysis under an extended threat model that includes malicious clients capable of intercepting communications.

    \item Full implementation and empirical evaluation of both approaches on the Flower framework with the PASTA/BFV HHE scheme, using the MNIST dataset. Results show that both mechanisms preserve model accuracy while adding negligible (masking) or modest (RSA encapsulation) overhead.
    
\end{enumerate}

The remainder of the paper is organized as follows: Section~\ref{sec:prelim} introduces background concepts on FL, HE, and HHE. Section~\ref{sec:related} surveys related work and highlights common challenges. Section~\ref{sec:proposed_sol} presents our proposed approaches and threat model. Section~\ref{sec:implementation_details} discusses implementation details, and Section~\ref{sec:experiments} reports our experimental results. Finally, Section~\ref{sec:conclusion} concludes with a summary and directions for future work.

\section{Preliminaries} \label{sec:prelim}
This section reviews background needed for our setting: centralized federated learning and aggregation, homomorphic encryption with a focus on BFV scheme, and HHE with HE-friendly ciphers (PASTA) and the homomorphic evaluation of symmetric decryption (HESD).

\subsection{Introduction to FL}

FL is a distributed machine learning paradigm that enables multiple devices or organizations to collaboratively train a shared global model without exposing their local data. FL can be \textit{centralized}, where a central server manages multiple clients and orchestrates the training process, or \textit{decentralized}, where no central server is required~\cite{liCentralizedDecentralizedFederated2025}. In this work we focus on the centralized paradigm, in which clients share model updates with the server rather than raw data~\cite{aledhariFederatedLearningSurvey2020}. 
Although raw data stays on client devices, updates can still leak private information through attacks such as gradient inversion, membership inference, and property inference. This motivates the integration of privacy-preserving techniques into FL.

Aggregation algorithms are central to FL: they combine local model updates into a global model, and the choice of algorithm impacts both model accuracy and computational overhead~\cite{electronics12102287}. The most common is \emph{Federated Averaging} (FedAvg)~\cite{li2024experimentalstudydifferentaggregation}: with client \(k \in \lbrace 1, \dots, K \rbrace\) holding \(n_k\) samples and sending local weights \(w_{t+1}^k\), the global update at round \(t+1\) is
\[
w_{t+1} = \sum_{k=1}^K \frac{n_k}{n} w_{t+1}^k, \quad \text{with } n=\sum_{k=1}^K n_k.
\]
\emph{Simple Averaging} (SimpAvg) is the unweighted counterpart. Besides FedAvg and its unweighted variant SimpAvg, other aggregation methods have been explored in the literature, such as FedSGD and dimension-reduction techniques~\cite{xuEdgeServerEnhanced2024,fontenla-romeroFedHEONNFederatedHomomorphically2023}, but these are outside the scope of our work, since in our evaluation we employ centralized FL with FedAvg.

FL deployments can be categorized as \emph{cross-silo}, involving a small number of institutional clients (e.g., hospitals), or \emph{cross-device}, involving many resource-constrained clients such as IoT devices that connect intermittently. We emulate the latter scenario, where communication bandwidth, client drop-outs, and limited computational resources are the main bottlenecks~\cite{kairouzAdvancesOpenProblems2021,majeedCrossSiloModelBasedSecure2021}. 

Despite the fact that raw data never leaves client devices, FL must still contend with both external (e.g., eavesdroppers) and internal threats (e.g., malicious participants). A common model is the \emph{honest-but-curious} server, which follows the protocol but attempts to infer private information from the updates it receives. Local model updates embed subtle patterns that may leak sensitive information~\cite{10089719,10.1007/978-981-97-5603-2_36,NIU2024404,Yu2023}, showing that the often-cited guarantee that ``no raw data leaves the device'' is insufficient for strong privacy.

\subsection{Homomorphic Encryption}

HE is a cryptographic technique that supports computation over encrypted data, resulting in ciphertexts that, when decrypted, produce the same result as if the operations had been applied to the plaintext~\cite{sathyaReviewHomomorphicEncryption2018}.
The type and number of operations that an HE scheme can support determine its classification. Namely, a \textit{Partially Homomorphic Encryption} (PHE) scheme allows a single type of arithmetic operation (e.g., addition or multiplication) an unlimited number of times; \textit{Somewhat Homomorphic Encryption} (SWHE) supports a limited set of operations, with restrictions on how many times they can be applied; and \textit{Fully Homomorphic Encryption} (FHE) enables multiple types of arithmetic operations to be performed an unlimited number of times.

Currently, there are four main FHE schemes: TFHE (Fast Fully Homomorphic Encryption over the Torus)~\cite{chillottiTFHEFastFully2018}, which operates at the bit level; BGV~\cite{brakerskiLeveledFullyHomomorphic} and BFV~\cite{fanSomewhatPracticalFully2012}, which are very similar and allow for exact computations over integers; and CKKS~\cite{cheonHomomorphicEncryptionArithmetic2016}, which supports approximate computations over complex numbers. 
However, since the security of HE schemes relies on complex mathematical problems, they tend to be {impractical for real-world use cases}, especially FHE schemes~\cite{gentryPaper}. Two of the main limitations of FHE are \textit{ciphertext expansion} and \textit{noise growth}~\cite{marcolla}. The former refers to the increase in ciphertext size when encrypting a plaintext or after every computation, and the latter refers to the accumulation of noise after each operation, especially costly ones such as multiplication. 
These limitations cause two major problems for implementation: ciphertext expansion makes communication, storage, and computation significantly more expensive, while noise growth limits the depth of computable circuits, since correct decryption cannot be ensured if the noise grows beyond a certain threshold~\cite{dobraunigPastaCaseHybrid2021}. 

To mitigate these drawbacks, new approaches such as HHE have been proposed. The basic idea consists of combining HE with SE, which is known for its fast encryption, efficient transmission, and constant ciphertext expansion.

\subsection{HE-friendly Ciphers}\label{background_HHE}

HHE is a cryptographic approach that combines symmetric ciphers with HE schemes to improve efficiency. In this setting, symmetric encryption is used for data encryption, while HE enables secure computations on the encrypted data. One notable example is the \textit{transciphering} framework~\cite{lauterCanHomomorphicEncryption2011}, which securely converts symmetric ciphertexts into homomorphic ciphertexts for further processing.
This hybrid approach significantly reduces the client-side computational burden, as the client does not need to homomorphically encrypt all of its data. It also minimizes communication overhead, since symmetric ciphertexts are more compact than homomorphic ones. These advantages make HHE especially attractive for IoT and other resource-constrained devices. However, they come at the cost of increased computational load on the server, which must handle the more demanding homomorphic evaluation.

To alleviate the server-side overhead, several symmetric ciphers have been specifically designed for compatibility with HE. These so-called HE-friendly ciphers reduce the cost of the \textit{Homomorphic Evaluation of Symmetric Decryption} (HESD) step, thereby improving the overall practicality of transciphering-based approaches. The most well-known HE-friendly ciphers include PASTA~\cite{dobraunigPastaCaseHybrid2021}, HERA~\cite{choTranscipheringFrameworkApproximate2020}, and Elisabeth~\cite{cosseronGloballyOptimizedHybrid2022}. PASTA is a stream cipher designed to minimize multiplicative depth, optimized for integer use cases over \(\mathbb{F}_p\) with \(p\) being a 16-bit prime. It is compatible with both the BGV and BFV schemes. HERA, in contrast, uses a randomized key schedule defined over \( \mathbb{Z}_q \), with \( q > 2^{16} \), and is primarily tailored for CKKS, though it remains compatible with BGV and BFV. Elisabeth targets TFHE and provides a wide range of operations for homomorphic evaluation on the server side. It is defined over \( \mathbb{Z}_q \) with \( q = 2^4 \).

The authors of PASTA released an open-source C++ framework\footnote{\url{https://github.com/isec-tugraz/hybrid-HE-framework}}, developed specifically for benchmarking HHE schemes. It includes implementations of several HE-friendly ciphers, such as PASTA, together with various HE schemes, including BGV via HElib~\cite{haleviDesignImplementationHElib2020}, BFV via Microsoft SEAL~\cite{dowlinManualUsingHomomorphic2017}, and TFHE using the TFHE library\footnote{\url{https://github.com/tfhe/tfhe}}.

\section{Related Work} \label{sec:related}
Several works have applied HE to enhance privacy and security in FL protocols. 
Some proposals rely on simplified schemes, namely PHE, as their main cryptographic primitive~\cite{rabieinejadTwoLevelPrivacyPreservingFramework2024,songSecureEfficientFederated2024,zhangPrivacyEAFLPrivacyEnhancedAggregation2023}. 
Other works employ FHE as the main component, typically within centralized topologies, since the substantial computational overhead of FHE requires a server to handle most of the processing. 
The advantage of this approach is flexibility: any aggregation algorithm can be used, as FHE supports arbitrary computations. 
For example, Ma et al.~\cite{maPrivacypreservingFederatedLearning2022}, Zhang et al.~\cite{zhangFaulttolerantFederatedLearning2024}, and Duy et al.~\cite{duyFedChainHunterReliablePrivacypreserving2023} apply FedAvg, while others have explored alternatives such as SimpAvg, FedSGD, or dimension-reduction techniques~\cite{tanPrivacypreservingFederatedLearning2024,fontenla-romeroFedHEONNFederatedHomomorphically2023,xuEdgeServerEnhanced2024}. 
However, even in centralized designs, the computational burden on clients remains high, which makes these systems impractical for resource-constrained environments.

To address these limitations, recent works have proposed using HHE instead of FHE in FL, thereby reducing the client-side overhead. 
The first HHE-based FL system was introduced by Correia et al.~\cite{correia2025}, who designed a centralized topology that combines BFV with the HE-friendly cipher PASTA. 
Their implementation, built on the PASTA framework\footnote{\url{https://github.com/isec-tugraz/hybrid-HE-framework}} and integrated into Flower~\cite{beutelFlowerFriendlyFederated2022}, showed that model performance can be maintained while significantly reducing client communication costs. 
However, server-side computation increased substantially due to the HESD step, and, more critically, all clients shared the same HE key pair, which weakens the threat model by enabling key misuse if one client acts maliciously.
For instance, a client with malicious intentions, Eve, intercepts the communication between another client, Alice, and the server. This exchange between Alice and the server are two ciphertexts: one PASTA ciphertext encrypting Alice’s model weights, and a BFV ciphertext encrypting Alice’s PASTA key.  Eve then attempts to access the secret information hidden in these ciphertexts. Even though she can not decrypt the PASTA ciphertext directly, she can use the BFV secret key (Eve and Alice have commom BFV keys) to decrypt the BFV ciphertext, obtaining Alice’s PASTA key. Following this, Eve can use this key to decrypt the PASTA ciphertext and retrieve Alice’s model weights. She can then apply any of the well known attacks for classic FL systems, such as inference attacks. 

At the same time, another work proposing a centralised HHE-FL system, that shares the above issue, was proposed by Nguyen et al.~\cite{michalas_FL}. However, in this work the authors use the RtF framework, which combines the Rubato cipher with FV and CKKS. In their design, model parameters are encrypted under Rubato, while the Rubato key is encrypted under FV by a Trusted Key Dealer (TKD), which partially addresses the key management challenge but places strong trust in the trusted key dealer. The server then performs HESD and aggregation (via SimpAvg) on the converted ciphertexts. Their results were consistent with Correia et al., showing preserved accuracy and reduced client communication, at the expense of higher server-side computation.

Our work addresses this open problem by eliminating the vulnerabilities associated with shared-key usage in HHE-FL. 
We integrate two alternative key-protection mechanisms into the workflow: \emph{masking}, where client keys are blinded before homomorphic encryption and later unblinded by the server, and \emph{RSA encapsulation}, where homomorphically encrypted keys are additionally wrapped under the server’s RSA public key. 
These approaches preserve the efficiency benefits of HHE while extending its security to adversarial settings with malicious participants.

\section{Threat Model and Proposed Approaches} \label{sec:proposed_sol}

In this section, we present our threat model and describe the proposed countermeasures that eliminate the shared-key liability in prior HHE-FL systems. 
We then introduce the overall architecture, provide a logical decomposition of the entities and their components, and describe the complete workflow of our approaches.

\subsection{Threat Model}\label{threat_model}

The security of the proposed approaches is based on the following assumptions about the setup and participants:  

\textit{A.1. Trusted setup phase:} All cryptographic keys, certificates, and auxiliary values (e.g., masks) are generated and distributed securely, with no adversarial interference.  

\textit{A.2. Clients are honest-but-curious but may maliciously intercept communications:} They follow the protocol correctly but may attempt to infer private information from intercepted messages.  

\textit{A.3. The server is honest-but-curious:} It executes the protocol correctly but may try to infer private information.  

\textit{A.4. No collusion occurs between the server and clients.}  

\textit{A.5. External adversaries may attempt to intercept communications but cannot compromise the underlying cryptographic primitives.}  

Under these assumptions, our system provides the following guarantees: 

\textit{G.1.} Confidentiality of individual client updates during transmission.  

\textit{G.2.} Confidentiality of individual client updates during aggregation at the server.  

\textit{G.3.} Confidentiality of the aggregated model weights.  

\textit{G.1} is ensured by client-side encryption: each client encrypts its local update with a symmetric key, and this key is further protected via masking or RSA encapsulation before being encrypted under the homomorphic public key. 
Thus, even if a client intercepts another’s message, the symmetric key cannot be misused. 
\textit{G.2} is ensured because the server applies HESD to convert symmetric ciphertexts into homomorphic ciphertexts, but never learns any plaintext update; without the homomorphic secret key, the server cannot decrypt intermediate results. 
Finally, \textit{G.3} follows from the fact that aggregation is performed directly over homomorphic ciphertexts, keeping the global model encrypted until decryption by authorized clients.

\paragraph{Comparison with Prior Threat Models.}  
The threat models of existing HHE-FL systems are weaker. 
Correia et al.~\cite{correia2025} assume honest clients and an honest-but-curious server, leaving the system vulnerable because all clients share the same HE key pair. 
Nguyen et al.~\cite{michalas_FL} adopt a similar semi-honest model but rely on a trusted key dealer to manage key encryption, introducing an additional trusted party. 
In contrast, our model explicitly considers malicious clients capable of intercepting traffic, and our countermeasures (masking and RSA encapsulation) eliminate this liability without introducing new trusted entities.

\subsection{System Overview}

\begin{figure}[ht]
    \centering
    \includegraphics[width=1\linewidth]{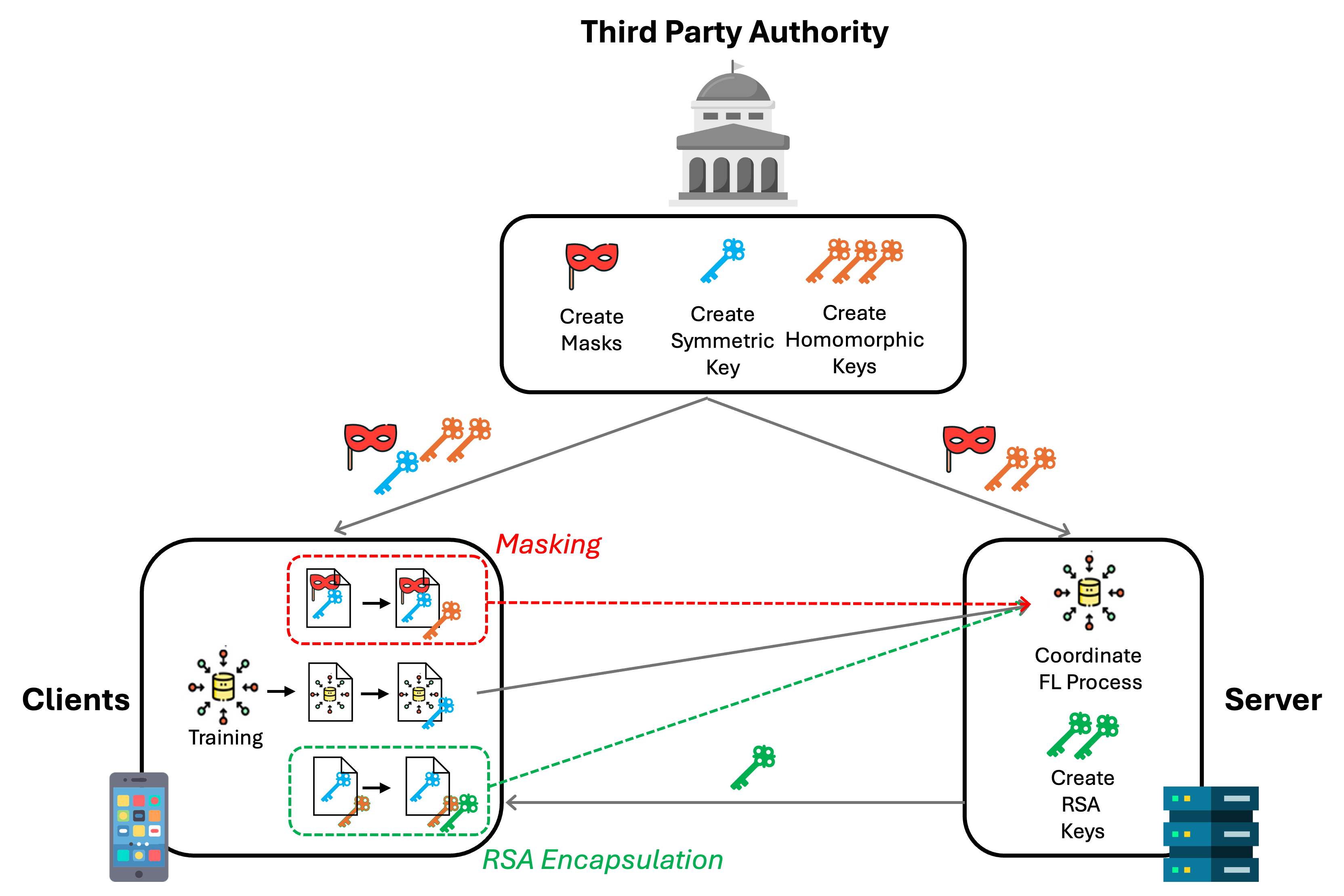}
    \caption{Architecture of the proposed approaches}
    \label{fig:hhe_with_eval}
\end{figure}

In this work, similarly to the work of Correia et al.~\cite{correia2025}, we consider three main entities: the \textit{Third-Party Authority} (TPA), the \textit{Server}, and the \textit{Clients}.
A high-level overview of the architecture is shown in Fig.~\ref{fig:hhe_with_eval}, showcasing all three entities, with only a single client for illustration purposes.

The TPA is fully trusted and is responsible for generating and securely distributing all PASTA keys and BFV key pair to the clients (one PASTA key for each client and one BFV key pair for all clients), as well as the masks, in the \textit{Masking} scenario, or the server’s RSA key pair, in the \textit{RSA Wrapping} scenario.
The \textit{Server} coordinates the FL training process, gathers encrypted updates from \textit{Clients}, performs secure aggregation, and distributes the updated global model parameters. 
In the \textit{RSA encapsulation} approach, the \textit{Server} also manages the RSA protocol.
The \textit{Clients} train their local models with private data, encrypt their local updates and symmetric key, send them to the \textit{Server}, and then decrypt the received global model parameters after aggregation.

\subsection{Logical View}

\begin{figure}[ht]
    \centering
    \includegraphics[width=1\linewidth]{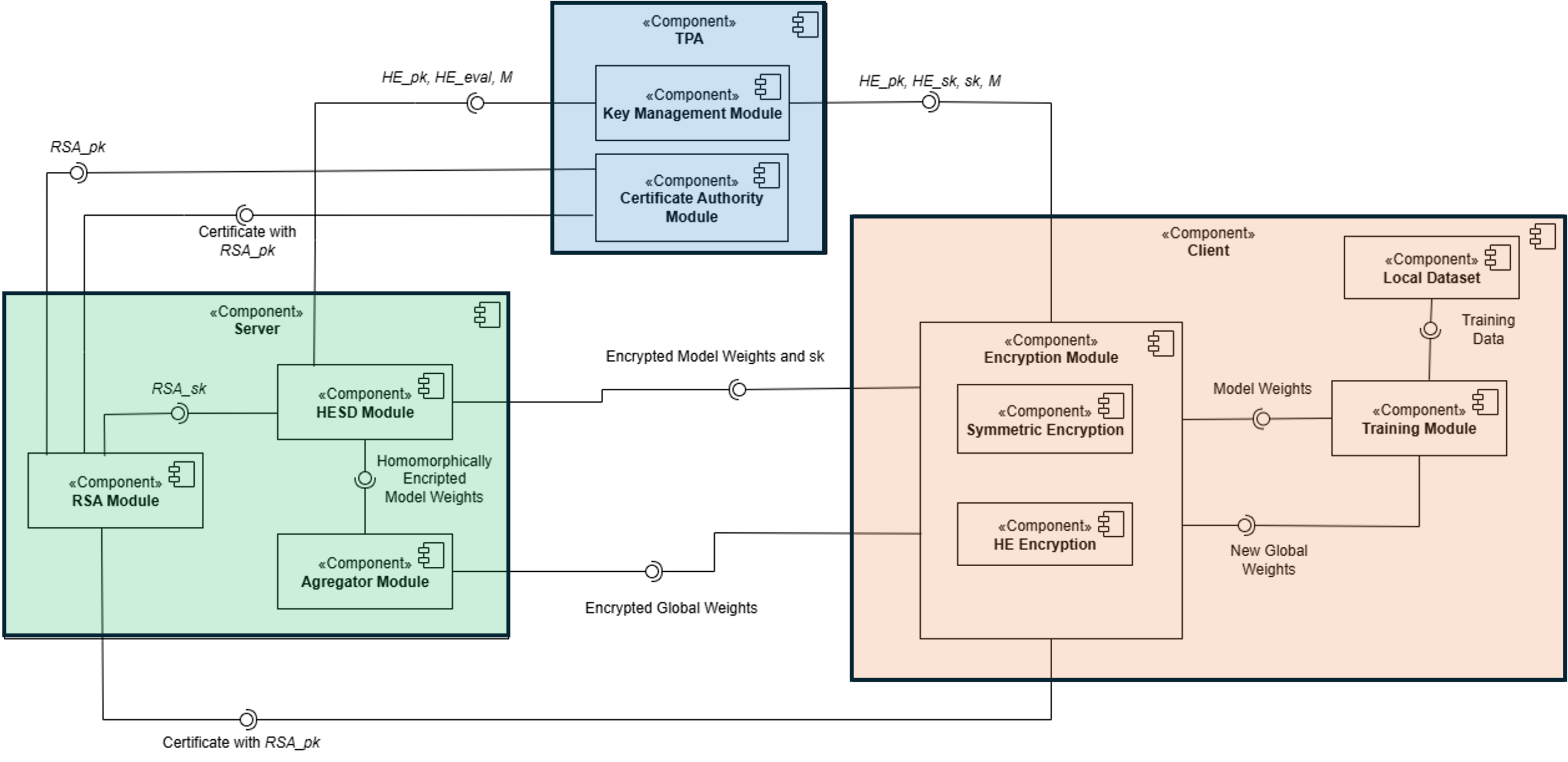}
    \caption{Logical View of the Proposed System}
    \label{fig:comp_diagram_mitigation}
\end{figure}

A logical view of the system is presented in Fig.~\ref{fig:comp_diagram_mitigation}.
This view decomposes the system into three main components, one for each entity. 
On the TPA side (blue in Fig.~\ref{fig:comp_diagram_mitigation}), the \textit{Key Management Module} is responsible for generating and securely distributing cryptographic keys to both client and server. In the \textit{Masking} approach, it is also responsible for creating a mask for each client, which is sent to the server and the respective client.  
On the client side (orange), the \textit{Training Module} trains the local model using data from the \textit{Local Dataset}. The \textit{Encryption Module} then encrypts the local model updates and the symmetric key, sending the encrypted data to the server. Once the updated global model is received, the client’s \textit{Encryption Module} decrypts it, and the \textit{Training Module} incorporates the global weights into the local model.  
On the server side (green), the \textit{HESD Module} performs HESD on the received model updates, generating homomorphic ciphertexts. These ciphertexts are then passed to the \textit{Aggregator Module}, which computes the new global model. The aggregated global model is then sent back to the clients.  
In the approach using \textit{RSA encapsulation}, the \textit{Certificate Authority Module} generates certificates containing the server’s RSA public key, ensuring its authenticity. On the server side, the \textit{RSA Module} manages the server’s RSA key pair and handles the distribution of the public key and its certificate to the clients.

\subsection{Workflow}
\begin{figure}
    \centering
    \includegraphics[width=0.6\linewidth]{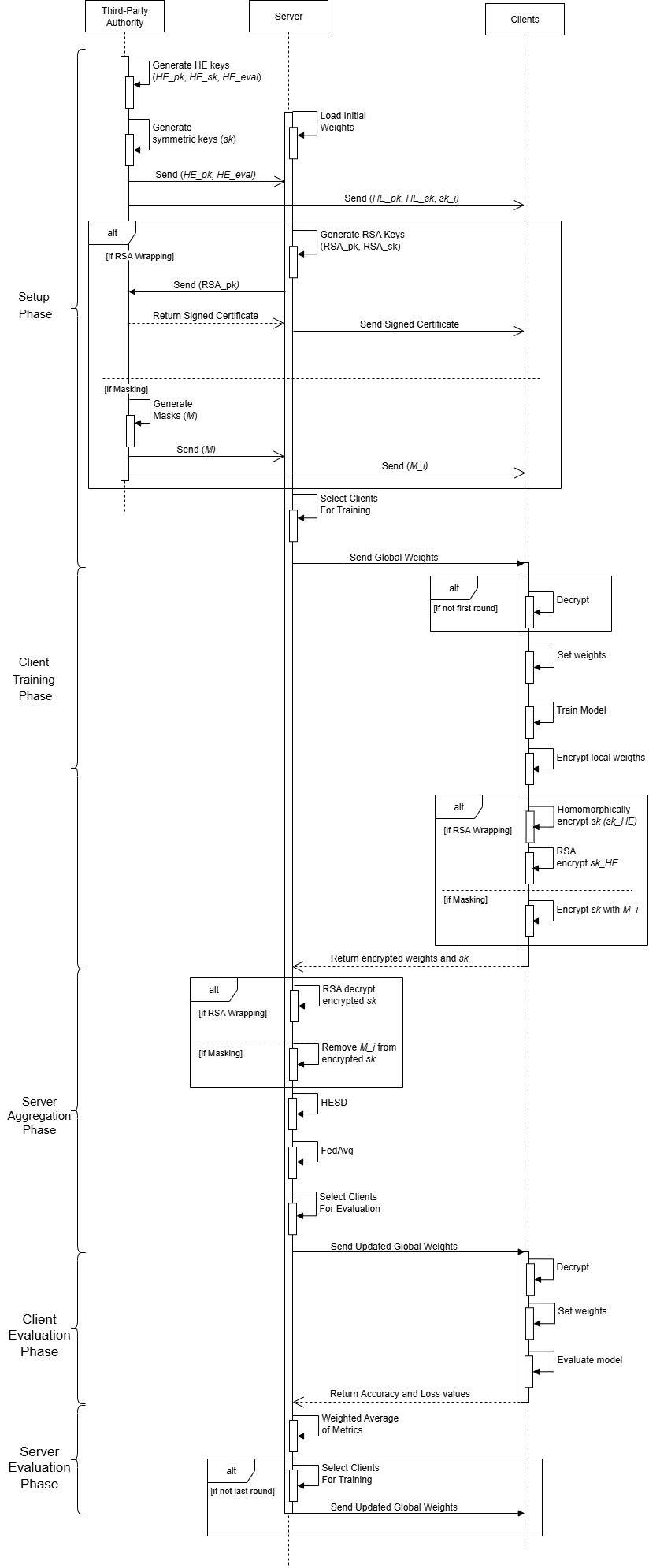}
    \caption{Workflow of the proposed approaches}
    \label{fig:sequence_diagram2}
\end{figure}

The workflow of our proposed approaches consists of five main phases: \textit{Setup Phase}, \textit{Client Training Phase}, \textit{Server Aggregation Phase}, \textit{Client Evaluation Phase}, and the \textit{Server Evaluation Phase}. The full workflow is depicted in Fig.~\ref{fig:sequence_diagram2}.

In the \textit{Setup Phase}, the TPA generates the set of homomorphic keys, (\textit{HE\_pk}, \textit{HE\_sk}, \textit{HE\_eval}) where \textit{HE\_pk} is the homomorphic public key, \textit{HE\_sk} is the homomorphic secret key, and \textit{HE\_eval} is the homomorphic evaluation key. Both \textit{HE\_eval} and \textit{HE\_pk} are sent to the server, while \textit{HE\_pk} and \textit{HE\_sk} are sent to the clients.
Additionally, the TPA generates a unique symmetric key \(sk_i\) for each client \(i\), where \(1 \le i \le N\) and $N\in \mathbb{N}$ denotes the total number of clients.
In the approach using \textit{Masking}, the TPA also generates a unique mask \(M_i\) for each client \(i\) and sends it to them. This mask is also shared with the server along with the corresponding client identifiers, enabling the server to associeate each client with its mask.
Alternatively, in the approach using \textit{RSA encapsulation}, the server is the one responsible for generating a single tuple of RSA keys, \((RSA\_pk, RSA\_sk\)), where \(RSA\_pk\) is the RSA public key and \(RSA\_sk\) is the corresponding secret key. To guarantee the authenticity of the server, the server requests the TPA to sign the \(RSA\_pk\). After receiving the signed certificate, the server shares it with all clients.
Then the server initializes a model and sends its weights to the clients, who use them as the starting point for local training.

After the initial phase, it follows the \textit{Client Training Phase}, where each client trains its local model, and then encrypts the obtained model weights \(w_i\) using \(sk_i\).
In the approach using \textit{Masking}, the clients compute 
\[
HE.Enc(sk + M).
\]
In the \textit{RSA encapsulation} approach the clients encrypt \(sk_i\) with \(HE\_pk\), producing \(sk^i_{HE}\), which is then encrypted with the server's \(RSA\_pk\) from the certificate, generating \(RSA.Enc(sk^i_{HE})\). 
Note that we first encrypt with HE and then encapsulate with RSA; RSA cannot handle such large plaintexts directly, so wrapping the (already homomorphically encrypted) key in RSA avoids plaintext size limits.
The resulting ciphertexts \({w^i_{SKE}}\) and, \(HE.Enc(sk + M)\) or \(RSA.Enc(sk^i_{HE})\), depending on the approach, together with the number of training samples  $$n_i = \frac{\textit{training\_data}_i}{batch\_size},$$ are sent to the server for aggregation.

In the \textit{Server Aggregation Phase}, the server has to first obtain \(sk^i_{HE}\) from either \(HE.Enc(sk + M)\) or \(RSA.Enc(sk^i_{HE})\). 
In the approach using \textit{Masking}, this is done by subtracting the masks \(M_i\) from the corresponding clients,
and in the other approach, the server applies the RSA decryption function \(RSA.Dec\) with the RSA secret key \(RSA\_sk\) on the RSA encrypted keys \(RSA.Enc(sk^i_{HE})\) to recover the homomorphically encrypted keys $sk^i_{HE}$.
Then, the server applies HESD on each \(w_{SKE}\) using the associated \(sk^i_{HE}\), obtaining \(w_{HE}\).
Once all updates are transformed, the server aggregates the homomorphic ciphertexts to compute the new global model weights. Finally, the server selects a new subset of clients and sends them the updated model for evaluation.

The fourth phase, \textit{Client Evaluation}, consists of selecting a subset of clients to receive the newly aggregated global model weights and decrypt them using \(HE\_sk\). They update their local models with these weights and evaluate them on their local test dataset. Each client then sends the resulting accuracy and loss metrics to the server, along with the number of test samples.

The \textit{Server Evaluation Phase} is the last phase in our workflow and it requires the server to aggregate the evaluation metrics received by the clients, using a chosen aggregation algorithm.

\paragraph{\textit{Masking} and \textit{RSA Wrapping} Correctness}
For the correctness of the \textit{Masking} approach, we need to ensure that when a client sends an homomorphic ciphertext encrypting the masked key, the server can retrieve the homomorphic ciphertext corresponding to the key, by removing the respective mask. More specifically, client $i$ sends
\[
HE.Enc(sk_i + M_i)
\]
to the server. Since the server knows the respective mask $M_i$, it computes
\[
HE.Enc(sk_i + M_i) - M_i = HE.Enc(sk_i),
\]
which is possible due to the homomorphic properties of $HE.Enc()$. This proves that the server can correctly retrieve $HE.Enc(sk_i)$ from $HE.Enc(sk_i + M_i)$.

For the correctness of the \textit{RSA Wrapping} approach, the scenario is similar, but instead we want to prove that the server can correctly unwrap the RSA ciphertext to obtain the homomorphic ciphertext of the symmetric key. In this scenario, client $i$ sends to the server
\[
RSA.Enc(sk^i_{HE}),
\]
which was encrypted using the server's RSA public key $RSA\_pk$. This means that the server can just use its RSA secret key, RSA\_sk, to decrypt this ciphertext as follows
\[
RSA.Dec(RSA.Enc(sk^i_{HE})) = sk^i_{HE} = HE.Enc(sk_i).
\]
This shows that the server can retrieve $HE.Enc(sk_i)$ correctly.

\paragraph{Discussion.}
Both proposed mechanisms strengthen HHE-FL against malicious clients, but with different trade-offs.  
Masking is lightweight and incurs negligible computational and communication overhead, but it requires careful management of masks by the server.  
RSA encapsulation builds on widely deployed public-key cryptography, reducing trust in mask management, but introduces additional cost for key encapsulation and decapsulation.  
Compared to Correia et al.~\cite{correia2025}, our approaches remove the unrealistic assumption that all clients are fully honest while still avoiding reliance on a trusted key dealer as in Nguyen et al.~\cite{michalas_FL}.  
This positions our work as a more practical solution for adversarial FL settings, particularly in cross-device deployments with resource-constrained clients.
\subsection{Security Analysis}

We now analyze the security of the proposed mechanisms under the threat model defined in Section~\ref{threat_model}. 
Our goal is to show that the confidentiality guarantees (G.1--G.3) are satisfied assuming the hardness of the underlying cryptographic primitives.

\paragraph{Masking-based protection.}
Each client $i$ transmits $(w^i_{SKE}, HE.Enc(sk_i + M_i))$, where $w^i_{SKE}$ are model parameters encrypted under the symmetric key $sk_i$, and $M_i$ is a random mask.  
Since $M_i$ is uniformly sampled and unknown to other clients, $HE.Enc(sk_i + M_i)$ is computationally indistinguishable from an encryption of a random value under the semantic security of HE.  
Consequently, even if another client intercepts this message, it cannot extract $sk_i$ or $w^i_{SKE}$ without the secret mask $M_i$ or the homomorphic secret key.  
At the server, the mask is subtracted in the homomorphic domain, yielding $HE.Enc(sk_i)$; security follows directly from the IND-CPA property of the BFV scheme.

\paragraph{RSA encapsulation.}
In this variant, each client sends $w^i_{SKE}$, together with $RSA.Enc(HE.Enc(sk_i))$.
The symmetric key is first encrypted under HE and then encapsulated with the server’s RSA public key.  
By the semantic security of RSA (under the RSA assumption in the random oracle model), only the server can recover $HE.Enc(sk_i)$.  
Therefore, even if another client intercepts the transmission, it cannot obtain $sk_i$ or $w^i_{SKE}$.  
This removes reliance on mask management while preserving confidentiality.

\paragraph{Aggregation and server-side confidentiality.}
For both approaches, the server converts symmetric ciphertexts into homomorphic ciphertexts via HESD using only $HE.Enc(sk_i)$.  
At no point does the server learn $sk_i$ or $w^i$ in plaintext.  
By the security of the BFV scheme, intermediate ciphertexts remain IND-CPA secure, and aggregation is performed entirely in the homomorphic domain.  
Thus, both individual client updates and the aggregated model remain confidential until decryption by authorized parties.

\paragraph{Comparison to prior work.}
Correia et al.~\cite{correia2025} and Nguyen et al.~\cite{michalas_FL} rely on a single global HE key pair across all clients, leaving them vulnerable to key misuse if one client intercepts another’s ciphertext.  
Our mechanisms eliminate this liability: masking relies on independent random masks, and RSA encapsulation ensures that only the server can access $HE.Enc(sk_i)$.  
Hence, our system achieves confidentiality against malicious clients, unlike prior HHE-FL systems.

\paragraph{Security claim.}
Formally, assuming (i) the IND-CPA security of the BFV scheme, (ii) the semantic security of the symmetric cipher (PASTA), and (iii) the RSA assumption, we claim that an adversary controlling a subset of clients and intercepting all network traffic cannot recover any other client’s plaintext updates with probability non-negligibly better than random guessing. 
Additionally, since all this primitives were chosen to ensure 128-bit security, we can assume that the overall system also ensures the 128-bit security.

\paragraph{Analysis on Combined Security}
In the proposed system, specifically in the \textit{RSA Wrapping} approach, we construct a ciphertext consisting of two layers of encryption: first a BFV encryption and then a RSA encryption, generating the ciphertext $RSA(BFV(m))$ encrypting $m$.
Since the BFV ciphertexts have to be split into chunks before being encrypted with RSA, some of its security properties might be lost. Therefore, we consider the security of these combined ciphertexts to be ensured by the security level of the RSA scheme, which is 128-bits.
We also discuss the unwrapping and the \textit{transciphering} steps to show that no leakage happens during these operations. 
The former consists of the transition
\[
RSA(BFV(m)) \rightarrow BFV(m)
\]
which happens by applying the RSA decryption algorithm using the server's RSA secret key. Due to the IND-CPA security of the BFV scheme, we know that the server can not distinguish the chunks from random noise, which ensures that there are no leakage.
The latter consists of the transition
\[
PASTA(m) \rightarrow BFV(m)
\]
which happens by homomorphically evaluating PASTA's decryption algorithm using $BFV(k)$, which decrypts the PASTA ciphertext within BFV's ciphertext space, where $k$ is PASTA's symmetric key.
The IND-CPA security of BFV ensures that ciphertexts resulting from the evaluation algorithm satisfy the indistinguishability condition, which ensures that  there is no leakage during the \textit{transciphering} step.

\section{Implementation Details} \label{sec:implementation_details}
In this section, we present the details behind the implementation of our approaches. 
We use, as a foundation, the implementation of Correia et al., which was made available to us by the authors. 
This implementation uses the Flower framework for FL and the combination of PASTA and BFV as the HHE scheme.
For the cryptographic layer, we rely on the code made available by the authors of PASTA\footnote{\url{https://github.com/isec-tugraz/hybrid-HE-framework}}, which is written in C++.
For the FL layer, both server and client components are implemented within Flower, which uses gRPC\footnote{\url{https://grpc.io/}} for communication. Each side also has specific subcomponents to coordinate the FL process: the server manages aggregation and evaluation, while clients manage local training and evaluation.
Since the PASTA/BFV framework is implemented in C++ and Flower is a Python-based framework, a \textit{pybind11}\footnote{\url{https://github.com/pybind/pybind11}} bridge was used to connect both layers.
RSA operations were implemented via the Python \textit{cryptography library}\footnote{\url{https://cryptography.io/en/latest/}}, operating outside the C++ cryptographic layer.

\paragraph{Implementation trade-offs.}
Some design choices were made to balance efficiency and practicality. 
First, because division is not natively supported in the ciphertext space of BFV, we delegate the final division by the global sample size to the clients, after decryption, rather than performing it at the server. 
Second, for the RSA encapsulation approach, the size of $sk_{HE}$ exceeds the plaintext capacity of RSA. To address this, we apply chunking: the encrypted key is split into segments that fit within the modulus size, encrypted individually, and later recombined at the server after decryption. 
While this introduces overhead, it is unavoidable for correctness and is evaluated experimentally in Section~\ref{sec:experiments}.

\paragraph{Security considerations.}
Cryptographic material such as symmetric keys, masks, and homomorphic keys is never exposed in plaintext beyond the client or TPA environment. 
The pybind11 bridge only exchanges serialized ciphertexts, never raw keys, thereby preserving security across the Python/C++ boundary. 
On the server side, masks are stored only in encrypted form, and RSA key pairs are generated and managed following the recommendations of NIST~\cite{barkerRecommendationKeyManagement2016}.

\paragraph{Reproducibility.}
Our implementation builds on publicly available open-source frameworks: the PASTA HHE library and the Flower FL framework. 
We extended these with modules for RSA encapsulation, masking, and key management. 
The code, including integration with pybind11 and experimental scripts, will be made available upon publication to facilitate reproducibility and comparison with related work.

\subsubsection{Server-Side Processing}

The server-side processing consists of first performing the HESD operations, which convert the PASTA ciphertexts into BFV ciphertexts, and then performing the \textit{FedAvg} aggregation of model updates.
Algorithm~\ref{alg:hesd_chunks} shows the chunk-based HESD procedure, in which each chunk of the symmetrically encrypted weights \(w_{SKE}\) is converted into a BFV homomorphic ciphertext \(w_{HE}\) before aggregation. To perform this transformation, the server first obtains the usable homomorphically encrypted symmetric key \(sk_{HE}\) for each client. 
In the \textit{RSA encapsulation} approach, the server decrypts the RSA-encrypted chunks of \(RSA.Enc(sk_{HE})\) and reconstructs the original \(sk_{HE}\). Alternatively, in the \textit{Masking} approach, the server subtracts the known mask \(M\) from the ciphertext \(HE.Enc(sk + M)\) to recover \(sk_{HE} = HE.Enc(sk)\).

\begin{algorithm}[!ht]
\caption{Chunk-Based HESD for a Single Client}
\label{alg:hesd_chunks}
\footnotesize
\begin{algorithmic}[1]
\Require Client's encrypted symmetric model parameters $w_{SKE}$, Client's encrypted symmetric key $sk_{HE}$, BFV cipher instance \texttt{bfv\_cipher}
\Require Protocol variant flag: \texttt{RSA} or \texttt{Masking}
\Require If \texttt{RSA}: RSA private key \(RSA\_sk\) 
\Require If \texttt{Masking}: Client's Mask $M$ 

\If{\texttt{RSA}}
    \State Decrypt $sk_{HE}$
    \State \quad \texttt{decrypted\_chunks} $\gets$ empty list
    \For{each \texttt{chunk} in $sk_{HE}$}
        \State  $ \texttt{dec\_chunk} \gets \texttt{RSA.Decrypt(chunk,\(RSA\_sk\))}$
        \State Append \texttt{dec\_chunk} to \texttt{decrypted\_chunks}
    \EndFor
    \State Reconstruct $sk_{HE}$
    \State \quad $sk_{HE}$ $\gets$ \texttt{concat(decrypted\_chunks)}

\ElsIf{\texttt{Masking}}
    \State Remove mask $M$ homomorphically
    \State \quad $sk_{HE} \gets sk_{HE} - M$
\EndIf

\State Switch to client's encrypted key
\State \quad \Call{\texttt{bfv\_cipher}.\texttt{set\_encrypted\_key}}{$sk_{HE}$} 

\State Perform HESD:
\State \quad \texttt{he\_params} $\gets$ empty list

\For{each \texttt{chunk} in $w_{SKE}$}
    \State  $ \texttt{he\_chunk} \gets \texttt{bfv\_cipher.HESD(chunk)}$
    \State Append \texttt{he\_chunk} to \texttt{he\_params}
\EndFor

\State \Return \texttt{he\_params}
\end{algorithmic}
\end{algorithm}

After HESD, the aggregation is performed using the FedAvg aggregation algorithm, which multiplies each weight vector by the client’s training sample count, and sums the results homomorphically. Since division is not naturally supported in the ciphertext space of BFV, in our approaches we delegate the division by the global total \(n\) to the clients.

\subsubsection{Client-Side Processing}

On each client device, processing begins with quantization and encryption of the local model updates using PASTA, and ends with decryption and de-quantization of the aggregated global model.
In our implementation, the quantization and encryption strategy, as well as the de-quantization and decryption strategy, follow the approach of Correia et al.~\cite{correia2025}.

Besides these two steps, each client is also required to securely prepare its symmetric key \(sk_i\) for transmission to the server. In the approach using \textit{RSA encapsulation}, the client first computes \(sk^i_{HE} = HE.Enc(sk_i)\), splits it into chunks to fit RSA encryption limits, and then encrypts each chunk using the server’s \(RSA\_pk\). Alternatively, in the approach using \textit{Masking}, the client computes \(HE.Enc(sk_i + M_i)\) and transmits this masked ciphertext to the server, as depicted in Algorithm~\ref{alg:sym_key_enc}.

\begin{algorithm}[ht]
\caption{Symmetric Key Encryption}
\label{alg:sym_key_enc}
\footnotesize
\begin{algorithmic}[1]
\scriptsize
\Require Client Symmetric key $sk_i$, Homomorphic public key $HE\_pk$, BFV cipher instance \texttt{BFV\_cipher}
\Require Protocol variant flag: \texttt{RSA} or \texttt{Masking}
\Require If \texttt{RSA}: RSA public key $RSA\_pk$, Max plaintext size \texttt{plaintext\_size} 
\Require If \texttt{Masking}: Client Mask $M_i$ 

\If{RSA}
    \State $enc\_key \gets BFV\_cipher.encrypt(sk_i)$
    \State Split $enc\_key$ into chunks
    \State \quad \texttt{chunked\_key} $\gets$ empty list
    \For{$i=0$ to length($enc\_key$) step \texttt{plaintext\_size}}
         \State \texttt{chunk} $\gets$ $enc\_key$$[i : i + \texttt{plaintext\_size}]$
        \State Append \texttt{chunk} to \texttt{chunked\_key}
    \EndFor
    
    \State Encrypt chunks with $RSA\_pk$
    \State \quad $sk^i_{HE}$ $\gets$ empty list
    \For{each $chunk$ in \texttt{chunked\_key}}
        \State $enc\_chunk \gets RSA.Enc(chunk, RSA\_pk)$
        \State Append $enc\_chunk$ to $sk^i_{HE}$
    \EndFor
\ElsIf{Masking}
    \State $sk^i_{HE} \gets BFV\_cipher.encrypt(sk_i + M_i)$
\EndIf

\State \Return $sk^i_{HE}$
\end{algorithmic}
\end{algorithm}

In summary, our implementation extends the PASTA/BFV HHE framework and the Flower FL framework with two new key-protection mechanisms: masking and RSA encapsulation. 
Both client and server roles were modified to integrate these mechanisms, while maintaining compatibility with the underlying training pipeline. 
In the next section, we evaluate these implementations experimentally, focusing on their accuracy, communication, and computation costs, and compare them against the baseline system of Correia et al.~\cite{correia2025}.

\section{Experimental Evaluation} \label{sec:experiments}

This section presents the experimental evaluation of our proposed approaches. 
We first describe the experimental setup, dataset, model, and configuration parameters, and then benchmark RSA performance to select suitable key sizes for the \textit{RSA encapsulation} variant. 
Next, we compare both of our mechanisms, \textit{Masking} and \textit{RSA encapsulation}, against the baseline system of Correia et al.~\cite{correia2025}. 
The evaluation covers model quality, communication cost, and computation cost, with results reported for both clients and server.

\subsection{Experimental Setup}

All experiments were conducted on Google Colab’s T4 High-RAM environment, which provides an Intel Xeon CPU @ 2.20GHz, an NVIDIA T4 GPU, 51GB of system RAM, and 15GB of GPU memory, offering hardware acceleration for training. 
The dataset used was MNIST~\cite{deng2012mnist}, a widely adopted benchmark for handwritten digit recognition. It contains 70,000 grayscale images of digits (0--9), with 60,000 for training and 10,000 for testing, each of size $28 \times 28$ pixels. 
The training dataset was partitioned into $N$ subsets using Flower’s \texttt{IidPartitioner}, with each local dataset further split into 80\% for training and 20\% for local testing. During training, 20\% of the local training set was reserved for validation. 
We used the same convolutional neural network (CNN) architecture as Correia et al.~\cite{correia2025}, based on the implementation by Sharma~\cite{sharma_mnist_cnn_kaggle}. The model consists of two convolutional layers, each followed by LeakyReLU activation and batch normalization, with max-pooling for downsampling. Feature maps are flattened, followed by dropout, and a final dense layer predicts one of 10 classes. The network has approximately 8,000 trainable parameters.

Table~\ref{tab:fl_parameters} summarizes the configuration used across all experiments. The same configuration was adopted in Correia et al.~\cite{correia2025}, enabling a fair comparison. 

\begin{table}[ht]
\scriptsize
\centering
\caption{Parameters used for the FL experiment}
\label{tab:fl_parameters}
\begin{tabular}{l l |l p{1.1cm}}
\toprule
Config. Name & Value & Config. Name & Value \\ \midrule
Clients & 12 & Rounds & 10 \\
Epochs & 10 & Classifier & CNN \\
Loss function & Categorical \linebreak Cross-entropy & Batch size & 64 \\
Optimizer & Nadam & Learning Rate & 0.001 \\
Clients per Training Phase & 4 & Clients per Evaluation Phase & 12 \\
Clip Range ($\alpha$) & 5 & Key size & 256 bits \\
Plaintext size & 128 bits & Ciphertext size & 128 bits \\
Plaintext Modulus & 65537 & Polynomial Degree Modulus & 16384 \\
Security level & 128-bit & & \\
\bottomrule
\end{tabular}
\end{table}
\paragraph{Parameter constraints.}
The HHE scheme used in this work operates in $\mathbb{Z}_q$ with $q = 2^{16} + 1$. 
As a result, server-side computations must remain within $[0, 2^{16} + 1)$. 
This constrains the product of quantized model weights, the number of participating clients, and the number of training batches per client to remain below $2^{16}$. 
Formally,
\[
\text{weights} \times \text{batches\_per\_client} \times \text{training\_clients} < 2^{16} + 1.
\]

\subsection{RSA Benchmarking}

In the RSA encapsulation approach, the homomorphically encrypted key $sk_{HE}$ must be chunked before RSA encryption due to plaintext size limits. 
To select suitable key sizes, we benchmarked RSA with 1024, 2048, 3072, and 4096-bit keys, measuring encryption and decryption time, ciphertext expansion, and number of chunks.

Results are reported in Table~\ref{tab:RSA_perf}. Larger key sizes reduce the number of chunks and ciphertext expansion but increase decryption cost. 
While RSA-1024 offers the fastest decryption, it is insecure for our setting and produces large ciphertext expansion. 
RSA-2048 offers moderate trade-offs but does not meet the 128-bit security level required to match PASTA. 
RSA-3072 and RSA-4096 both achieve adequate security, with RSA-3072 balancing ciphertext size and runtime better, while RSA-4096 slightly reduces communication overhead at the cost of higher decryption latency.

\begin{table}[ht]
\small
\centering
\caption{Average Performance and Output Size Comparison Across RSA Key Sizes (Ptx - Plaintext ; Ctx - Ciphertext)}
\begin{tabular}{cccccccc}

\toprule
\makecell{RSA \\ Key size \\ (bits)} & \makecell{Enc. \\ (s)} & \makecell{Dec. \\ (s)} & \makecell{Ptx Size \\ (bytes)} & \makecell{Max \\ Ptx Size \\ (bytes)} & \makecell{Total \\ Chunks} & \makecell{Ctx Size \\ (bytes)} & \makecell{Ctx Size \\ (MB)} \\ \midrule
1024 & 0.51 & 3.3 & 1826326 & 62 & 29456 & 4742416 & 4.52 \\ 
2048 & 0.43 & 7.91 & 1826049 & 190 & 9611 & 2777579 & 2.65 \\ 
3072 & 0.31 & 11.74 & 1826287 & 318 & 5743 & 2394831 & 2.28 \\
4096 & 0.35 & 18.72 & 1826243 & 446 & 4095 & 2231775 & 2.13 \\ \bottomrule
\end{tabular}
\label{tab:RSA_perf}
\end{table}
Based on these results, we focus our evaluation of the RSA encapsulation approach on RSA-3072 and RSA-4096, as they provide the necessary 128-bit security level.

\subsection{Results}
The results of the analysis are presented below, highlighting the key findings and their implications.
\paragraph{Model quality.}

\begin{figure}[ht]
    \centering
    \begin{minipage}[b]{0.45\linewidth}
        \includegraphics[width=\linewidth]{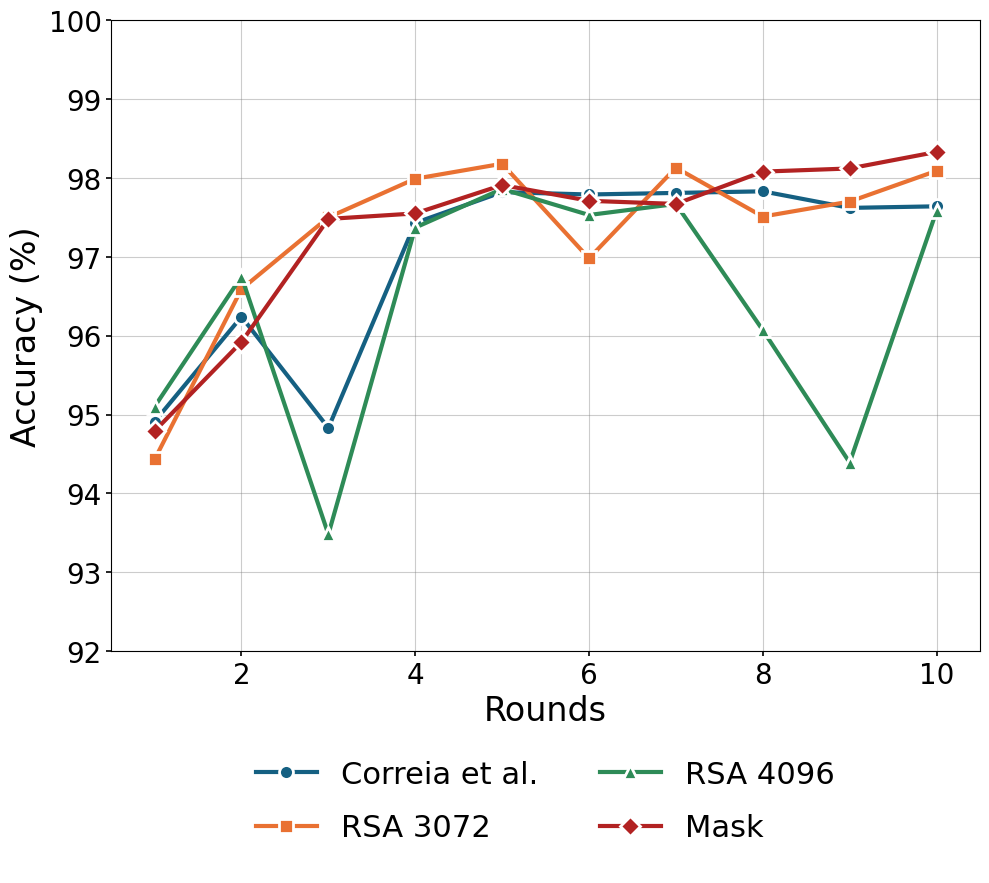}
    \end{minipage}
    \hfill
    \begin{minipage}[b]{0.45\linewidth}
        \includegraphics[width=\linewidth]{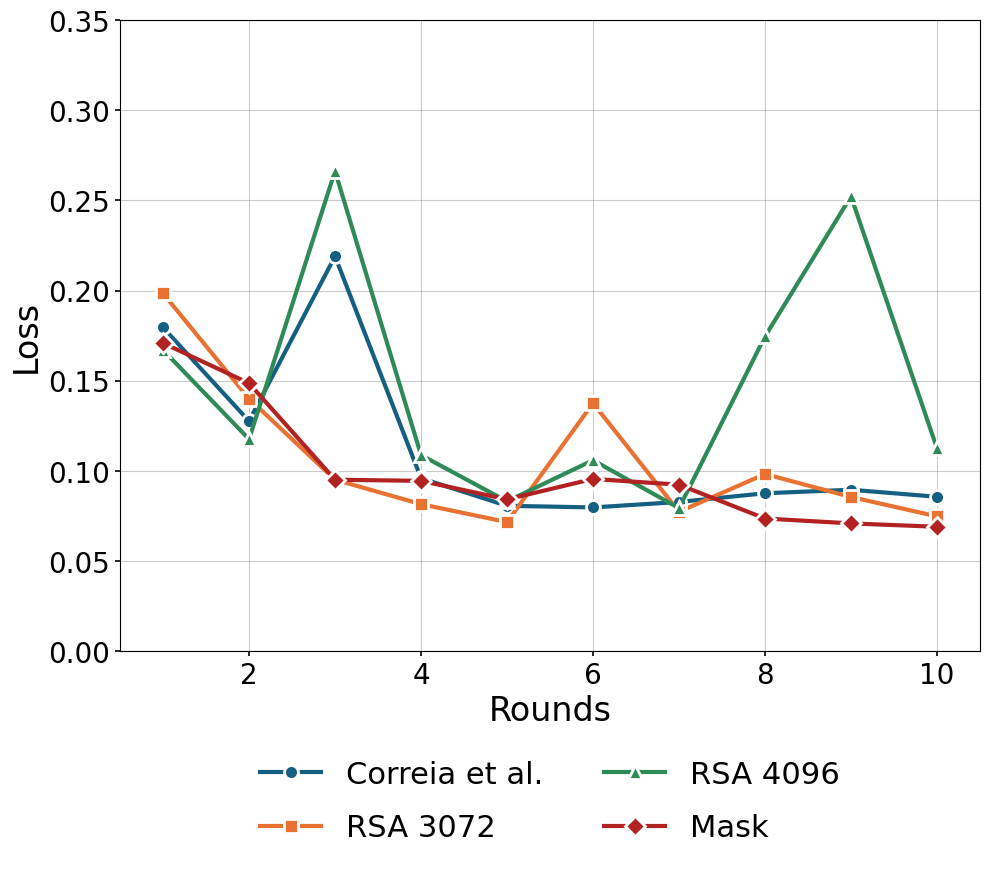}
    \end{minipage}
    \caption{Training Performance Across Rounds: (a) Accuracy; (b) Loss}
    \label{fig:accuracy_loss_fl}
    \vspace{-0.2cm}
\end{figure}

Fig.~\ref{fig:accuracy_loss_fl} shows accuracy and loss over training rounds. 
Our approaches closely match or even slightly outperform the baseline of Correia et al.~\cite{correia2025}. 
Final results are summarized in Table~\ref{tab:global_model_values}. 
Masking achieves the highest accuracy (98.35\%), while RSA-4096 incurs only a minor drop (97.28\%). 
All approaches preserve model quality effectively.

\begin{table}[ht]
\centering
\caption{Global Model Comparisons}
\scriptsize
\begin{tabular}{lccccc}
\toprule
\multicolumn{1}{c}{} & Accuracy(\%) & Loss & Precision(\%) & Recall(\%) & F1-score(\%) \\ \midrule
Correia et al. & 97.39 & 0.0962 & 97.50 & 97.39 & 97.39 \\ 
RSA 3072 & 98.04 & 0.0737 & 98.09 & 98.04 & 98.05 \\ 
RSA 4096 & 97.28 & 0.1152 & 97.38 & 97.28 & 97.28 \\ 
Masking & 98.35  & 0.0622 & 98.37 & 98.35 & 98.35 \\ \bottomrule
\end{tabular}

\label{tab:global_model_values}
\end{table}

\paragraph{Communication costs.}
For clients, masking incurs the same ciphertext size as Correia et al.~\cite{correia2025} ($\approx$1.79 MB per client). 
RSA encapsulation increases communication by 1.31$\times$ (RSA-3072) and 1.22$\times$ (RSA-4096). 
On the server side, communication overhead is identical across all approaches since aggregation occurs in BFV ciphertext space.

\paragraph{Computation costs (clients).}
Table~\ref{tab:client_computation} reports client runtimes. 
Masking adds negligible overhead ($\approx$0.018s, performed at initialization). 
RSA encapsulation adds 0.35--0.38s for key chunking and encryption. 
Overall, client runtimes remain in the 11.9--12.4s range, only slightly higher than the baseline.

\
\begin{table}[ht]
\centering
\caption{Average Client Computation Cost}

\scriptsize
\begin{tabular}{lcccc}
\toprule
Operation (s) & Correia et al. & RSA-3072 & RSA-4096 & Masking \\ \midrule
De-serialization  & 0.42480 & 0.48112 & 0.47115 & 0.44298 \\ 
Decryption & 0.42206 & 0.49451 & 0.48567 & 0.46385  \\  
Training  & 10.70805 & 10.70805 & 10.70805 & 10.70805 \\  
Evaluation  & 1.07516 & 1.07516 & 1.07516 & 1.07516 \\ 
Weights Encryption  & 0.32867 & 0.33840447 & 0.33226 & 0.33888 \\ 
RSA Key Encryption  & - & 0.35177 & 0.38373 & - \\  
Key Masking\textsuperscript{a} & - & - & - &  0.01845 \\ 
Serialization & 0.00128 & 0.00240 & 0.00212 & 0.00123 \\ \midrule
Training runtime & 11.88486 & 12.37625 & 12.38298 & 11.95498 \\ 
Evaluation runtime & 1.92330 & 2.05319 & 2.03410 & 1.98321  \\ \bottomrule
\end{tabular}
\label{tab:client_computation}
\\
\textsuperscript{a} The time reported for Key Masking is not included in the phases runtime since it can be performed during client initialization.
\end{table}

\paragraph{Computation costs (server).}
Server runtime is dominated by HESD, which is common to all approaches ($\approx$1400s per client). 
RSA encapsulation adds 12s (RSA-3072) or 19s (RSA-4096) for decryption per client, while masking adds only 0.001s. 
Table~\ref{tab:server_computation} summarizes the results. 
Although RSA introduces measurable overhead, it remains insignificant compared to HESD.

\begin{table}[ht]
\centering
\caption{Average Server Aggregation Phase Computation Cost Comparison}
\scriptsize
\begin{tabular}{lcccc}
\toprule
Operation (s) & Correia et al. & RSA-3072 & RSA-4096 & Masking \\ \midrule
De-serialization  & 0.0009 & 0.0009 & 0.0008 & 0.0010 \\ 
HESD per client  & 1451.0 & 1431.5 & 1439.8 & 1400.0 \\ 
RSA Key Decrypt per client  & - & 12.0 & 19.3 & - \\ 
Unmasking per client & - & - & - & 0.0014 \\ 
Aggregation  & 0.2174 & 0.2222 & 0.2244 & 0.2212 \\ 
Serialization  & 0.3629 & 0.3626 & 0.3620 & 0.3528 \\ \midrule
Total runtime & 5804.5803 & 5774.5857 & 5836.9872 & 5600.5806 \\ \bottomrule
\end{tabular}
\label{tab:server_computation}
\end{table}
\subsection{Discussion}

The results show that both mechanisms preserve accuracy while significantly improving the threat model. 
Masking is extremely lightweight, with negligible communication and computation overhead. 
RSA encapsulation provides stronger guarantees without reliance on mask management but at the cost of moderate runtime and communication overhead. 
Both approaches outperform prior HHE-FL systems in terms of security while maintaining practicality for cross-device federated learning with resource-constrained clients.

Regarding scalability, all the bottlenecks are expected to appear on the server side, since it has to manage more clients, whereas the client protocols can happen simultaneously. 
The main issue will be on the HESD step, which is by far the most expensive operation performed on the server side, as the experimental results showed. 
However, regarding the proposed techniques, we expect the \textit{Masking} approach to outperform the \textit{RSA Wrapping} approach, since it is a much more lightweight solution, taking only 0.0014s per client to remove the masks on the server side.
Both RSA-3072 and RSA-4096 introduce a runtime cost of 12s and 19.3s per client, respectively, to perform RSA decryption, which is 8500$\times$ and 13500$\times$ more than the masking approach, making them obviously worse choices in settings with a large number of clients, since these costs scale proportionally to the number of clients.


\section{Conclusion} \label{sec:conclusion}
In this work, we revisited the security assumptions of existing HHE-based federated learning approaches and showed that the common practice of sharing a single homomorphic key pair across all clients leaves prior systems exposed to malicious participants. To address this, we integrated two alternative key-protection mechanisms, \emph{masking} and \emph{RSA encapsulation}, into the HHE-FL workflow. Both mechanisms prevent key misuse by other clients while preserving the efficiency advantages of HHE.

We implemented both variants on top of Flower using the PASTA/BFV stack and evaluated them on MNIST with 12 clients. The results indicate that model quality is maintained (up to 98.35\% accuracy with masking) while overheads remain modest: masking incurs negligible runtime and communication cost; RSA-3072/4096 increases client communication by only $1.31\times/1.22\times$ and adds a small client-side time penalty ($\approx 1.04$--$1.05\times$). On the server, the dominant cost continues to be HESD (shared by all systems), with RSA decapsulation adding a comparatively minor overhead.

Overall, our study demonstrates that it is possible to harden HHE-FL against malicious clients without sacrificing practicality for cross-device deployments. Future work includes (i) exploring lightweight multi-key HE or key-homomorphism to remove residual trust in a single HE secret key, (ii) dropout-robust training via secret sharing or coded computation, and (iii) parallel and hardware-assisted HESD to reduce the server bottleneck.

\begin{credits}
\subsubsection{\ackname}
This work has been supported by the PC2phish project, which has received funding from FCT with Refª: 2024.07648.IACDC. 
Furthermore, this work also received funding from the project
UID/00760/2025.

\subsubsection{\discintname}
\end{credits}
%
%
%
\bibliographystyle{splncs04}
\bibliography{refs}
%




\end{document}